\documentstyle[12pt,amstex]{article} 

  \newtheorem{df}{Definition}[section]

  \newtheorem{prop}[df]{Proposition}
  \newtheorem{cor}[df]{Corollary}
  
\begin{document}
 \title{ Nonlinear Sigma Models in $(1+2)$-Dimensions \\
        and \\ An Infinite Number of Conserved Currents}
 \author{ 
  Kazuyuki FUJII\thanks{Department of Mathematics, 
  Yokohama City University, 
  Yokohama 236, 
  Japan, \endgraf 
  {\it E-mail address}: fujii{\char'100}yokohama-cu.ac.jp} \ and 
  Tatsuo SUZUKI\thanks{Department of Mathematics, 
  Waseda University, 
  Tokyo 169, 
  Japan, \endgraf 
  {\it E-mail address}: 695m5050{\char'100}mn.waseda.ac.jp}}
 \date{}
 \maketitle
\begin{abstract}
We treat in this paper non-linear sigma models such as $CP^1$-model, $QP^1$-model and etc, in $1+2$ dimensions. For submodels of such ones we definitely construct an infinite number of nontrivial conserved currents. Our result is a generalization of that of authors (Alvarez, Ferreira and Guillen). 
\end{abstract}
 \section{Introduction}
Integrable models in $(1+1)$-dimensions are good toy models to understand $(1+3)$-dimensional field theories such as Yang-Mills one or Yang-Mills-Higgs one. See, for example, \cite{Zak}.

They have an infinite number of nontrivial conserved currents, which usually correspond to an infinite dimensional Lie algebra such as Kac-Moody algebra, Virasoro algebra or W-algebra. See \cite{Kac}.

For many such models the equations of motion are formulated in terms of the zero-curvature condition $F_{01}=0$. Under this condition a holonomy operator doesn't depend on paths whose end points are kept fixed. Then we can construct the conserved charges from this operator. See \cite{A-F-G}. This method is very powerful and useful. But this is typical of one or two-dimensional cases. To generalize the method to higher dimensional cases beyond the two dimensions is not easy task. But recently Alvarez, Ferreira and Guillen in the interesting paper \cite{A-F-G} proposed a new idea to generalize the method in two dimensions. In particular they defined a three dimensional integrability and applied their method to the $CP^1$-model in $(1+2)$-dimensions to obtain an infinite number of nontrivial conserved currents. But their results (calculation) are not complete. The aim of this paper is to give explicit forms to conserved currents of a submodel of $CP^1$-model and also apply their !
method to other non-linear sigma models in $(1+2)$-dimensions to obtain an infinite number of nontrivial conserved currents. 

 \section{Review of [3]}
We first make a review of \cite{A-F-G} within our necessity. 

Let $M$ be a $(1+d)$-dimensional differential manifold and $x_0 \in M$ a fixed point. We denote by $G$ a Lie group and by $L_0 M$ a path space on $M$ starting from $x_0$,
\begin{equation}
 L_0 M \equiv \{ x:[0,2\pi] \rightarrow M | x(\sigma) \in M, x(0)=x_0 \}.
\end{equation}
Let $W$ be a holonomy operator $W:L_0 M \rightarrow G$. That is, $W$ is defined by the differential equation
\begin{equation}
 \frac{dW}{d\sigma}+A_{\mu}\frac{dx^{\mu}}{d\sigma}W=0 \quad 
  \mbox{on} \quad \Gamma \in L_0 M,
 \label{eqn:2}
\end{equation}
where $A_{\mu}(\mu =0,1,\cdots ,d)$ is a connection on a principal $G$-bundle on $M$. We set the initial condition $W(0)=I$. We note that $W$ is not a local function of $x=(x^{\mu})$ and is usually given by the path-ordered integral. 

We investigate the conditions under which $W$ becomes local. For this aim we study a variation of $W$ under deformations of $\Gamma$ keeping the initial and end points (the boundary). The result is 
\begin{equation}
 W^{-1}(2\pi)\delta W(2\pi)
  =\int_0^{2\pi} d\sigma 
     W^{-1}F_{\mu \nu} W \frac{dx^{\mu}}{d\sigma}\delta x^{\nu},
 \label{eqn:3}
\end{equation}
where $F_{\mu \nu}$ is the curvature of $A_{\mu}$, 
\begin{equation}
 F_{\mu \nu} \equiv \partial_{\mu} A_{\nu}-\partial_{\nu} A_{\mu}
                       +[A_{\mu},A_{\nu}].
\end{equation}
If this curvature vanishes
\begin{equation}
 F_{\mu \nu}=0,
 \label{eqn:5}
\end{equation}
then $\delta W(2\pi)=0$ from (\ref{eqn:3}), so $W$ becomes path-independent. Namely $W$ is a local function on $\Gamma \quad (W=W(x^{\mu}(\sigma)))$.

From (\ref{eqn:2}) $A_{\mu}$ is written as 
\begin{equation}
 A_{\mu}=-\partial_{\mu}W W^{-1}.
\end{equation}
$A_{\mu}$ is a pure gauge.

Next we consider a two-dimensional ``holonomy" operator. Let $\Gamma$ be a fixed loop at $x_0$. We denote by $S_{\Gamma}M$ a space of surfaces on $M$ with the boundary $\Gamma$,
\begin{equation}
 S_{\Gamma}M \equiv \{ \Sigma : \mbox{a surface on $M$}
                        |\partial \Sigma =\Gamma \}.
\end{equation}
$\Gamma$ is parametrized by $\sigma \in [0,2\pi ]$, so we parametrized $\Sigma$ as follows.

We scan $\Sigma$ with loops passing through $x_0$ and being parametrized by $\tau \in [0,2\pi ]$ such that $\tau =0$ is the infinitesimal loop around $x_0$ and $\tau =2\pi $ is $\Gamma$. We want to identify a surface $\Sigma$ in $S_{\Gamma}M$ with such a parametrization by $\tau \in [0,2\pi ]$. Here we introduce a gauge field $A_{\mu}$ and an anti-symmetric tensor field $B_{\mu \nu}$. Then a two-dimensional ``holonomy" operator $V:S_{\Gamma}M \rightarrow G$ is defined by the differential equation
\begin{equation}
 \frac{dV}{d\tau}-VT(B,A;\tau)=0 \quad \mbox{on} \quad \Sigma \in S_{\Gamma}M 
 \label{eqn:8}
\end{equation}
where
\begin{equation}
 T(B,A;\tau) \equiv \int_0^{2\pi}d\sigma W^{-1}B_{\mu \nu}W
      \frac{\partial x^{\mu}}{\partial \sigma}
      \frac{\partial x^{\nu}}{\partial \tau}
     =\frac12
      \int_0^{2\pi}d\sigma W^{-1}B_{\mu \nu}W
      \frac{\partial (x^{\mu}, x^{\nu})}{\partial (\sigma, \tau)}
 \label{eqn:9}
\end{equation}
and $W$ is given by (\ref{eqn:2}) and $x^{\mu}=x^{\mu}(\sigma,\tau)$. We set the initial condition $V(0)=I$. In general there are an infinite number of methods to scan $\Sigma$ in $S_{\Gamma}M$. The quantity $V$ in (\ref{eqn:8}) should not depend on such methods. Therefore we assume $A_{\mu}$ be flat $(F_{\mu \nu}=0)$. Then $W$ becomes local by the arguments below (\ref{eqn:5}), so $T=T(B,A;\tau)$ in (\ref{eqn:9}) becomes also local. But $V$ itself is still non-local. We research the conditions under which $V$ becomes local. For this aim we study a variation of $V$ under deformations of $\Sigma$ keeping the boundary $\Gamma$. The result is 
\begin{eqnarray}
 && \hspace{-2cm}
    \delta V(2\pi)V^{-1}(2\pi) \nonumber\\
  =\int_0^{2\pi} & d\tau & V(\tau)
    \left\{
     \int_0^{2\pi} d\sigma
      W^{-1}(D_{\lambda}B_{\mu \nu}+D_{\mu}B_{\nu \lambda}+D_{\nu}B_{\lambda \mu}) W \right. \nonumber\\
 && \hspace{1cm} \times 
      \frac{\partial x^{\mu}}{\partial \sigma}
      \frac{\partial x^{\nu}}{\partial \tau} \delta x^{\lambda} \nonumber\\
 && \left.
    -[T(B,A;\tau), \int_0^{2\pi} d\sigma
      W^{-1}B_{\mu \nu}W \frac{dx^{\mu}}{d\sigma}\delta x^{\nu}]
    \right\} V^{-1}(\tau), 
 \label{eqn:10}
\end{eqnarray}
where
$$
 D_{\lambda}B_{\mu \nu} \equiv
  \partial_{\lambda}B_{\mu \nu}+[A_{\lambda}, B_{\mu \nu}]. 
$$
If the R.H.S. of (\ref{eqn:10}) vanishes, then $\delta V(2\pi)=0$. Namely $V$ becomes surface-independent or $V$ becomes a local function on $\Sigma$. 

A comment is in order. The R.H.S. of (\ref{eqn:10}) is identified with the curvature of a principal $G$-bundle on the loop space $\Omega G$. See the appendix in \cite{A-F-G}. Now we can construct conserved charges from $V$. This is the main story of \cite{A-F-G}.

Next we must study the vanishing conditions of the R.H.S. of (\ref{eqn:10}). First we restrict a Lie group $G$. Let $\frak g$ be its Lie algebra. Here we assume $\frak g$ is non-semisimple. If we set $\frak p$ be the radical of $\frak g$, then $\frak g$ is decomposed into $\frak g=\frak h \oplus \frak p$ by the Levi's theorem \cite{Jac}. Now we assume
\begin{equation}
 A_{\mu} \in \frak g \quad \mbox{and} \quad \mbox{$B$}_{\mu \nu} \in \frak p.
 \label{eqn:11}
\end{equation}
Since $\frak p$ is the radical (abelian) and $W^{-1}B_{\mu \nu}W \in \frak p$, so we have
\begin{equation}
 [T(B,A;\tau), \int_0^{2\pi} d\sigma
   W^{-1}B_{\mu \nu}W \frac{dx^{\mu}}{d\sigma}\delta x^{\nu}]=0.
\end{equation}
Second we restrict gauge fields $A_{\mu}$ and $B_{\mu \nu}$. We have choosed $A_{\mu}$ a flat connection $(F_{\mu \nu}=0)$. Therefore if we have 
\begin{equation}
 G_{\mu \nu \lambda} \equiv 
  D_{\lambda}B_{\mu \nu}+D_{\mu}B_{\nu \lambda}+D_{\nu}B_{\lambda \mu}=0,
\end{equation}
the R.H.S. of (\ref{eqn:10}) vanishes. That is to say, we call
\begin{equation}
 F_{\mu \nu}=0 \quad \mbox{and} \quad G_{\mu \nu \lambda}=0
 \label{eqn:14}
\end{equation}
local integrability conditions under (\ref{eqn:11}). 

Up to now our arguments are based on any $(1+d)$-dimensional manifold $M$. From here let us restrict to $(1+2)$-dimensional case. For $B_{\mu \nu}$ we consider its dual
\begin{equation}
 \tilde{B}^{\mu} \equiv \frac12 \epsilon^{\mu \nu \lambda}B_{\nu \lambda},
 \label{eqn:15}
\end{equation}
where $\epsilon^{012}=1=-\epsilon_{012}$. Then we have $G_{\mu \nu \lambda}=0 \Leftrightarrow D_{\mu}\tilde{B}^{\mu}=0$. Therefore we can write local integrability conditions (\ref{eqn:14}) as 
\begin{equation}
 F_{\mu \nu}=0 \quad \mbox{and} \quad D_{\mu}\tilde{B}^{\mu}=0.
 \label{eqn:16}
\end{equation}
This is a typical feature of 3-dimensions. 

Next let us construct a non-semisimple Lie algebra $\hat{\frak g}$ starting from $\frak g$. Let $R$ be a representation of $\frak g$, $R:\frak g \rightarrow \frak g\frak l(\mbox{$P$})$, where $P$ is a representation space (abelian ideal). The construction of $\hat{\frak g}$ is as follows:
\begin{equation}
 0 \rightarrow P \rightarrow \hat{\frak g} \rightarrow \frak g \rightarrow 
 \mbox{0}.
 \label{eqn:17}
\end{equation}
Let $\{ T_a \}$ be a basis of $\frak g$ and $\{ P_i \}$ of $P$. The commutation relations in $\hat{\frak g}$ are
\begin{eqnarray}
 &&[T_a, T_b]=f_{ab}^c T_c, \nonumber\\
 &&[T_a, P_i]=P_j R_{ji}(T_a), \label{eqn:18}\\
 &&[ P_i, P_j ]=0. \nonumber
\end{eqnarray}
Now we choose $A_{\mu}$ and $B_{\mu \nu}$ as 
\begin{equation}
 A_{\mu} \in \frak g \quad \mbox{and} \quad \mbox{$B$}_{\mu \nu} \in \mbox{$P$}
\end{equation}
satisfying (\ref{eqn:16}). Then the current 
\begin{equation}
 J_{\mu} \equiv W^{-1}\tilde{B}_{\mu}W
 \label{eqn:21}
\end{equation}
is conserved 
\begin{equation}
 \partial_{\mu}J^{\mu}=W^{-1}D_{\mu}\tilde{B}^{\mu}W=0.
\end{equation}
On the other hand since $J_{\mu} \in P$
\begin{equation}
 J_{\mu}=\sum_{i=1}^{\mbox{\footnotesize{dim$P$}}} J_{\mu}^i P_i,
\end{equation}
so $\{ J_{\mu}^i | 1 \le i \le \mbox{dim}P \} $ is a set of conserved currents. This is just the one which we are looking for. Therefore if the number of different representations $R$ is infinite, we can get an infinite number of conserved currents in this way.
 \section{$CP^1$ model and its submodel}\label{section:2}
In this section we consider the $CP^1$-model in $(1+2)$-dimensions as an effective example of the preceding theory. $CP^1$ (1-dimensional complex projective space) is identified with $SU(2)/U(1)$ and the embedding $i:CP^1 \rightarrow SU(2)$ is 
\begin{equation}
 i(CP^1)=\left\{
          \frac{1}{\sqrt{1+|v|^2}}
           \left(
            \begin{array}{cc}
               1     & v \\
            -\bar{v} & 1 \\
            \end{array}
           \right)
          |v \in \bf C
         \right\}
\end{equation}
well-known. But according to \cite{A-F-G}, we set $v=iu$ $(u \in \bf C)$ to obtain 
\begin{equation}
 i(CP^1)=\left\{
          \frac{1}{\sqrt{1+|u|^2}}
           \left(
            \begin{array}{cc}
               1     & iu \\
            i\bar{u} & 1 \\
            \end{array}
           \right)
          |u \in \bf C
         \right\} .
 \label{eqn:25}
\end{equation}
This form becomes useful later on. We note here that $CP^1$ is identified with the projection space
\begin{equation}
 CP^1 \cong \left\{
             \frac{1}{1+|u|^2}
           \left(
            \begin{array}{cc}
               1     & -iu   \\
            i\bar{u} & |u|^2 \\
            \end{array}
           \right)
          |u \in \bf C
         \right\} .
\end{equation}
The action of $CP^1$-model in $(1+2)$-dimensions is given by
\begin{equation}
 \mbox{$\cal{A}$}(u) \equiv \int d^3 x \frac{\partial^{\mu}\bar{u}\partial_{\mu}u}
                             {(1+|u|^2)^2},
\end{equation}
where \quad $u:M^{1+2} \rightarrow \bf C$. Its equation of motion is 
\begin{equation}
 (1+|u|^2)\partial^{\mu}\partial_{\mu}u
  -2\bar{u}\partial^{\mu}u\partial_{\mu}u =0.
 \label{eqn:28}
\end{equation}
This model is invariant under the transformation 
\begin{equation}
 u \rightarrow \frac{1}{u}.
 \label{eqn:29}
\end{equation}
It is well-known that this model has three conserved currents
\begin{eqnarray}
 && J_{\mu}^{Noet}=\frac{1}{(1+|u|^2)^2}
       (\partial_{\mu}u \bar{u}-u\partial_{\mu}\bar{u}), \label{eqn:30}\\
 && j_{\mu}=\frac{1}{(1+|u|^2)^2}
       (\partial_{\mu}u +u^2 \partial_{\mu}\bar{u}), \label{eqn:31}\\
 && \mbox{and the complex conjugate} \quad \bar{j_{\mu}},
  \label{eqn:32}
\end{eqnarray}
corresponding to the number of generators of $SU(2)$. 

Next applying the preceding theory to this case, we must identify a Lie algebra $\hat{\frak g}$ in (\ref{eqn:17}). Let $\frak g$ be $\frak s\frak l(\mbox{2,\bf C})$, the Lie algebra of $SL(2,\bf C)$. Let $\{ T_{+}, T_{-}, T_3 \} $ be generators of $\frak s\frak l(\mbox{2,\bf C})$ satisfying 
\begin{equation}
 [T_3,T_{+}]=T_{+}, \quad [T_3,T_{-}]=-T_{-}, \quad [T_{+},T_{-}]=2T_3.
\end{equation}
Usually we choose
\begin{equation}
 T_{+}= \left(
          \begin{array}{cc}
            0 & 1 \\
            0 & 0 
          \end{array}
        \right), \quad 
 T_{-}= \left(
          \begin{array}{cc}
            0 & 0 \\
            1 & 0 
          \end{array}
        \right), \quad 
 T_3  = \frac12
        \left(
          \begin{array}{cc}
            1 &  0 \\
            0 & -1 
          \end{array}
        \right) .
\end{equation}
From here we consider a spin $j$ representation of $\frak s\frak l(\mbox{2,\bf C})$. Then $\hat{\frak g}$ in (\ref{eqn:18}) is given by 
\begin{eqnarray}
 && [T_3, P_m^{(j)}] = mP_m^{(j)}, \nonumber\\
 && [T_{\pm}, P_m^{(j)}] = \sqrt{j(j+1)-m(m \pm 1)}P_{m \pm 1}^{(j)}, \\
 && [P_m^{(j)}, P_n^{(j)}] = 0 \nonumber
\end{eqnarray}
where $m \in \{ -j,-j+1, \cdots, j-1,j \} $ and $\{ P_m^{(j)}|-j \le m \le j \}$ is a set of generators of the representation space $P \cong {\bf C}^{2j+1}$. We note that $P_j^{(j)}$ $(P_{-j}^{(j)})$ is the highest (lowest) spin state.
Now we must choose gauge fields $A_{\mu}$, $B_{\mu \nu}$ or $A_{\mu}$, $\tilde{B}_{\mu}$ in (\ref{eqn:15}) to satisfy (\ref{eqn:16}). From (\ref{eqn:25}), we set
\begin{equation}
 W \equiv W(u)=
          \frac{1}{\sqrt{1+|u|^2}}
           \left(
            \begin{array}{cc}
               1     & iu \\
            i\bar{u} & 1 \\
            \end{array}
           \right)
 \label{eqn:36}
\end{equation}
and choose
\begin{eqnarray}
 A_{\mu} & \equiv & -\partial_{\mu}W W^{-1} \nonumber\\
         & = & \frac{-1}{1+|u|^2}
               \{ i\partial_{\mu}u T_{+}+i\partial_{\mu}\bar{u} T_{-}+
                  (\partial_{\mu}u \bar{u}-u\partial_{\mu}\bar{u})T_3 \}, 
 \label{eqn:37}\\
 \tilde{B}_{\mu}^{(1)} & \equiv & \frac{1}{1+|u|^2}
             (\partial_{\mu}u P_1^{(1)}-\partial_{\mu}\bar{u} P_{-1}^{(1)}).
 \label{eqn:38}
\end{eqnarray}
From this choice we have $F_{\mu \nu}=0$. A comment is in order. The Gauss decomposition of $W$ in (\ref{eqn:36}) is given by
\begin{equation}
 W=W_1 \equiv e^{iuT_{+}}e^{\varphi T_3}e^{i\bar{u}T_{-}}
\end{equation}
or
\begin{equation}
 W=W_2 \equiv e^{i\bar{u}T_{-}}e^{-\varphi T_3}e^{iuT_{+}}
\end{equation}
where $\varphi = \log{(1+|u|^2)}$. Now it is easy to show that $CP^1$-model satisfies the local integrability conditions 
\begin{equation}
 F_{\mu \nu}=0 \quad \mbox{and} \quad D_{\mu}\tilde{B}^{\mu (1)}=0 
\end{equation}
from (\ref{eqn:37}) and (\ref{eqn:38}). Therefore the conserved currents (\ref{eqn:21}) are
\begin{equation}
 J_{\mu}^{(1)} \equiv W^{-1} \tilde{B}_{\mu}^{(1)} W 
               = j_{\mu}P_1^{(1)}-\sqrt{2}iJ_{\mu}^{Noet}P_0^{(1)}
                         -\bar{j}_{\mu}P_{-1}^{(1)}, 
\end{equation}
where coefficients are given by (\ref{eqn:30}), (\ref{eqn:31}), (\ref{eqn:32}). This is not the end of our story. Moreover we consider more extended situation in (\ref{eqn:37}), (\ref{eqn:38}). That is, we choose
\begin{equation}
 \tilde{B}_{\mu}^{(j)} \equiv \frac{1}{1+|u|^2}
             (\partial_{\mu}u P_1^{(j)}-\partial_{\mu}\bar{u} P_{-1}^{(j)})
\end{equation}
instead of $\tilde{B}_{\mu}^{(1)}$ in (\ref{eqn:38}). In this case $P_1^{(j)}$ $(P_{-1}^{(j)})$ is not the highest (lowest) spin state. If we assume $D_{\mu}\tilde{B}^{\mu (j)}=0$, where
\begin{eqnarray}
 D_{\mu}\tilde{B}^{\mu (j)} &=& \frac{1}{(1+|u|^2)^2}
     \left\{ \sqrt{j(j+1)-2} \ 
      (-i\partial_{\mu}u\partial^{\mu}u P_2^{(j)}
       +i\partial_{\mu}\bar{u}\partial^{\mu}\bar{u} P_{-2}^{(j)}) 
     \right. \nonumber\\
  && \hspace{1cm}
     +\{(1+|u|^2)\partial^{\mu}\partial_{\mu}u
        -2\bar{u}\partial^{\mu}u\partial_{\mu}u \} P_1^{(j)} \nonumber\\
  && \hspace{1cm}
     \left.
     +\{(1+|u|^2)\partial^{\mu}\partial_{\mu}\bar{u}
        -2u\partial^{\mu}\bar{u}\partial_{\mu}\bar{u} \} P_{-1}^{(j)}
     \right\} ,
\end{eqnarray}
we must add a new constraint in addition to the equation of motion (\ref{eqn:28}):
$$ (1+|u|^2)\partial^{\mu}\partial_{\mu}u
        -2\bar{u}\partial^{\mu}u\partial_{\mu}u =0 \quad 
\mbox{and} \quad
 \partial^{\mu}u\partial_{\mu}u=0. $$
Namely
\begin{equation}
 \partial^{\mu}\partial_{\mu}u=0 \quad \mbox{and} \quad 
 \partial^{\mu}u\partial_{\mu}u=0. 
\end{equation}
We call this one a submodel of $CP^1$-model according to \cite{A-F-G}. Then the conserved currents are 
\begin{equation}
 J_{\mu}^{(j)} \equiv W^{-1} \tilde{B}_{\mu}^{(j)} W
   =\sum_{m=-j}^{j} J_{\mu}^{(j,m)}P_m^{(j)}. 
\end{equation}
In \cite{A-F-G} they determined $\{ J_{\mu}^{(j,m)} | \ |m| \le j \}$ for $j=1,2,3$ only and left the remaining cases. In fact to determine these for any $j \in N$ is not so easy (hard work). But we did this. Namely the result is
\begin{prop}\label{prop:2.1} we have \\
(a) for $j \geq m \geq 1$, 
 \begin{eqnarray}
   J_{\mu}^{(j,m)} &=&
    \sqrt{\frac{(j+m)!}{j(j+1)(j-m)!}}
    \frac{(-iu)^{m-1}}{(1+|u|^2)^{j+1}} \nonumber\\
  && \hspace{1mm} \times
   \left(  
    \sum_{n=0}^{j-m} \alpha _n^{(j,m)} |u|^{2n} \partial_{\mu} u + (-1)^{j-m}
    \sum_{n=0}^{j-m} \alpha _{j-m-n}^{(j,m)} |u|^{2n} u^2 \partial_{\mu}\bar{u}
   \right) ,
 \end{eqnarray}
(b) for $m=0$, 
 \begin{equation}
   J_{\mu}^{(j,0)}=
    -i \sqrt{j(j+1)}
    \frac{(\bar{u} \partial_{\mu} u - u \partial_{\mu} \bar{u})}
         {(1+|u|^2)^{j+1}} 
    \sum_{n=0}^{j-1} \gamma_n^{(j,0)} |u|^{2n} , 
 \end{equation}
(c) for $j \geq m \geq 1$, 
\begin{equation}
  J_{\mu}^{(j,-m)}=(-1)^m  J_{\mu}^{(j,m)^{\dag}}, 
\end{equation}
where coefficients are
\begin{eqnarray}
 \alpha _n^{(j,m)} &=&
   (-1)^n
   \frac{n!}{(m+n-1)!} 
    \left(
     \begin{array}{c}
      j-m \\
       n  
     \end{array}
    \right)
    \left(
     \begin{array}{c}
       j+1 \\
        n  
     \end{array}
    \right) , \\
   \gamma_n^{(j,0)} &=&
   (-1)^n
   \frac{1}{j} 
    \left(
     \begin{array}{c}
       j \\
       n  
     \end{array}
    \right)
    \left(
     \begin{array}{c}
        j \\
       n+1  
     \end{array}
    \right) .
\end{eqnarray}
\end{prop}
 \section{$CP^1$-like models and its submodels}
In this section, we consider $CP^1$-like sigma models in $(1+2)$-dimensions. First of all, we fix $j \in N$. The action of such a model is given by
\begin{equation}
 \mbox{$\cal{A}$}_j(u) \equiv \int d^3 x \frac{\partial^{\mu}\bar{u}\partial_{\mu}u}
                               {(1+|u|^2)^{j+1}},
 \label{eqn:52}
\end{equation}
where $u: M^{1+2} \rightarrow \bf C$. 
When $j=1$, (\ref{eqn:52}) reduces to $CP^1$-model. A comment is now in order. The action (\ref{eqn:52}) is not invariant under the transformation $u \rightarrow 1/u$ in (\ref{eqn:29}), so we may consider an invariant action
\begin{equation}
 \mbox{$\cal{\tilde{A}}$}_j(u) \equiv \int d^3 x 
     \frac{(1+|u|^{2(j-1)})\partial^{\mu}\bar{u}\partial_{\mu}u}
                               {(1+|u|^2)^{j+1}}.
\end{equation}
But for the sake of simplicity, we consider (\ref{eqn:52}) only in this paper. 
The equation of motion of (\ref{eqn:52}) reads 
\begin{equation}
 (1+|u|^2)\partial^{\mu}\partial_{\mu}u
        -(j+1)\bar{u}\partial^{\mu}u\partial_{\mu}u =0. 
\end{equation}
Taking an analogy of section \ref{section:2}, we set $A_{\mu}$ the same as (\ref{eqn:37}) and $\tilde{B}_{\mu}$ as
\begin{equation}
 \tilde{B}_{\mu} = \frac{1}{1+|u|^2}
             (\partial_{\mu}u P_j^{(j)}-\partial_{\mu}\bar{u} P_{-j}^{(j)})
\end{equation}
where $P_j^{(j)}$ $(P_{-j}^{(j)})$ is the highest (lowest) spin state. Let us calculate $D_{\mu}\tilde{B}^{\mu}$. 
\begin{eqnarray}
 D_{\mu}\tilde{B}^{\mu} &=& 
    -i\sqrt{2j}
      \frac{\partial^{\mu}\bar{u}\partial_{\mu}u}{(1+|u|^2)^2}P_{j-1}^{(j)}
    +i\sqrt{2j}
      \frac{\partial^{\mu}\bar{u}\partial_{\mu}u}{(1+|u|^2)^2}P_{-j+1}^{(j)}
       \nonumber\\
 && +\frac{(1+|u|^2)\partial^{\mu}\partial_{\mu}u
        -(j+1)\bar{u}\partial^{\mu}u\partial_{\mu}u
        +(j-1)u\partial^{\mu}\bar{u}\partial_{\mu}u}
          {(1+|u|^2)^2}P_{-j}^{(j)} \nonumber\\
 && +\frac{(1+|u|^2)\partial^{\mu}\partial_{\mu}\bar{u}
        -(j+1)u\partial^{\mu}\bar{u}\partial_{\mu}\bar{u}
        +(j-1)\bar{u}\partial^{\mu}\bar{u}\partial_{\mu}u}
          {(1+|u|^2)^2}P_j^{(j)} .
\end{eqnarray}
If we assume $D_{\mu}\tilde{B}^{\mu}=0$, then we have
\begin{equation}
  (1+|u|^2)\partial^{\mu}\partial_{\mu}u
        -(j+1)\bar{u}\partial^{\mu}u\partial_{\mu}u=0
  \quad \mbox{and} \quad \partial^{\mu}\bar{u}\partial_{\mu}u=0. 
\end{equation}
We again call this a submodel of $CP^1$-like model. For this model, the local integrability conditions $( F_{\mu \nu}=0 \ \mbox{and} \ D_{\mu}\tilde{B}^{\mu}=0 )$ are satisfied. Therefore, the conserved currents are
\begin{equation}
 J_{\mu} \equiv W^{-1} \tilde{B}_{\mu} W
   =\sum_{m=-j}^{j} J_{\mu}^{(j,m)}P_m^{(j)}. 
\end{equation}
We can determine $J_{\mu}^{(j,m)}$ completely. 
\begin{prop}For $j \geq m \geq -j$, we have
 \begin{equation}
  J_{\mu}^{(j,m)}=\alpha \frac{\bar{u}^{j-m}\partial_{\mu}u}{(1+|u|^2)^{j+1}}
                  +\beta \frac{u^{j+m}\partial_{\mu}\bar{u}}{(1+|u|^2)^{j+1}},
 \end{equation}
where $\alpha$ and $\beta$ are some constant (we don't need the explicit forms).
\end{prop}
Analysing this proposition, we can remove the constraint $j \geq |m|$. Namely, we have
\begin{cor}
 \begin{equation}
  J_{\mu}^{(n)}=\frac{\bar{u}^n \partial_{\mu}u}{(1+|u|^2)^{j+1}},
   \quad n \in Z
 \end{equation}
and its complex conjugate $\bar{J}_{\mu}^{(n)}$ are conserved currents. 
\end{cor}
 \section{$QP^1$ model and its submodel}
In this section, we consider the $QP^1$-model in $(1+2)$-dimensions. $QP^1$ (1-dimensional quasi projective space) is identified with $SU(1,1)/U(1)  \cong D=\{z \in {\bf C}| \ |z|<1 \} $ and the embedding $i:QP^1 \rightarrow SU(1,1)$ is 
\begin{equation}
 i(QP^1)=\left\{
          \frac{1}{\sqrt{1-|u|^2}}
           \left(
            \begin{array}{cc}
               1     & iu \\
           -i\bar{u} & 1 \\
            \end{array}
           \right)
          |u \in D
         \right\} .
\end{equation}
Here $D$ is the Poincare disk. We note here that $QP^1$ is identified with the quasi projection space 
\begin{equation}
 QP^1 \cong \left\{
             \frac{1}{1-|u|^2}
           \left(
            \begin{array}{cc}
               1     & -iu   \\
           -i\bar{u} &-|u|^2 \\
            \end{array}
           \right)
          |u \in D
         \right\} .
\end{equation}          
See \cite{Fuj}, in detail. 

The action of $QP^1$-model in $(1+2)$-dimensions is given by 
\begin{equation}
 \mbox{$\cal{A}$}(u) \equiv \int d^3 x \frac{\partial^{\mu}\bar{u}\partial_{\mu}u}
                             {(1-|u|^2)^2},
\end{equation}
where \quad $u:M^{1+2} \rightarrow D$. Its equation of motion is 
\begin{equation}
 (1-|u|^2)\partial^{\mu}\partial_{\mu}u
  +2\bar{u}\partial^{\mu}u\partial_{\mu}u =0.
\end{equation}
This model is invariant under the transformation $ u \rightarrow 1/u
$ in (\ref{eqn:29}). This model has three conserved currents
\begin{eqnarray}
 && J_{\mu}^{Noet}=\frac{1}{(1-|u|^2)^2}
       (\partial_{\mu}u \bar{u}-u\partial_{\mu}\bar{u}), \label{eqn:65}\\
 && j_{\mu}=\frac{1}{(1-|u|^2)^2}
       (\partial_{\mu}u -u^2 \partial_{\mu}\bar{u}), \label{eqn:66}\\
 && \mbox{and the complex conjugate} \quad \bar{j_{\mu}},
 \label{eqn:67}
\end{eqnarray}
corresponding to the number of generators of $SU(1,1)$. The complexification of both $SU(2)$ in section \ref{section:2} and $SU(1,1)$ in this section is just $SL(2,\bf C)$. Therefore, the arguments in section \ref{section:2} are still valid in this section. Namely we set
\begin{equation}
 W \equiv W(u)=
          \frac{1}{\sqrt{1-|u|^2}}
           \left(
            \begin{array}{cc}
               1     & iu \\
           -i\bar{u} & 1 \\
            \end{array}
           \right) .
\end{equation} 
For this, the Gauss decomposition is given by
\begin{equation}
 W=W_1 \equiv e^{iuT_{+}}e^{\varphi T_3}e^{-i\bar{u}T_{-}}
\end{equation}
or
\begin{equation}
 W=W_2 \equiv e^{-i\bar{u}T_{-}}e^{-\varphi T_3}e^{iuT_{+}}
\end{equation}
where $\varphi = \log{(1-|u|^2)}$. We choose $A_{\mu}$ and $\tilde{B}_{\mu}^{(1)}$ as 
\begin{eqnarray}
 A_{\mu} & \equiv & -\partial_{\mu}W W^{-1} \nonumber\\
         & = & \frac{1}{1+|u|^2}
               \{ -i\partial_{\mu}u T_{+}+i\partial_{\mu}\bar{u} T_{-}+
                  (\partial_{\mu}u \bar{u}-u\partial_{\mu}\bar{u})T_3 \}, \\
 \tilde{B}_{\mu}^{(1)} & \equiv & \frac{1}{1-|u|^2}
             (\partial_{\mu}u P_1^{(1)}+\partial_{\mu}\bar{u} P_{-1}^{(1)}).
 \label{eqn:72}
\end{eqnarray}
Then, we easily have 
\begin{equation}
 F_{\mu \nu}=0 \quad \mbox{and} \quad D_{\mu}\tilde{B}^{\mu (1)}=0, 
\end{equation}
so the conserved currents are
\begin{equation}
 J_{\mu}^{(1)} \equiv W^{-1} \tilde{B}_{\mu}^{(1)} W 
               = j_{\mu}P_1^{(1)}+\sqrt{2}iJ_{\mu}^{Noet}P_0^{(1)}
                         +\bar{j}_{\mu}P_{-1}^{(1)}, 
\end{equation}
where coefficients are (\ref{eqn:65}), (\ref{eqn:66}), (\ref{eqn:67}). 

Next we consider the extended situation as shown in section \ref{section:2}. We choose
\begin{equation}
 \tilde{B}_{\mu}^{(j)} \equiv \frac{1}{1-|u|^2}
             (\partial_{\mu}u P_1^{(j)}+\partial_{\mu}\bar{u} P_{-1}^{(j)})
\end{equation}
instead of $\tilde{B}_{\mu}^{(1)}$ in (\ref{eqn:72}), and calculate $D_{\mu}\tilde{B}_{\mu}^{(j)}$. 
\begin{eqnarray}
 D_{\mu}\tilde{B}^{\mu (j)} &=& \frac{1}{(1-|u|^2)^2}
     \left\{ \sqrt{j(j+1)-2} \ 
      (-i\partial_{\mu}u\partial^{\mu}u P_2^{(j)}
       +i\partial_{\mu}\bar{u}\partial^{\mu}\bar{u} P_{-2}^{(j)}) 
     \right. \nonumber\\
  && \hspace{1cm}
     +\{(1-|u|^2)\partial^{\mu}\partial_{\mu}u
        +2\bar{u}\partial^{\mu}u\partial_{\mu}u \} P_1^{(j)} \nonumber\\
  && \hspace{1cm}
     \left.
     +\{(1-|u|^2)\partial^{\mu}\partial_{\mu}\bar{u}
        +2u\partial^{\mu}\bar{u}\partial_{\mu}\bar{u} \} P_{-1}^{(j)}
     \right\} .
\end{eqnarray}
Therefore, we consider a submodel of $QP^1$-model, 
\begin{equation}
 \partial^{\mu}\partial_{\mu}u=0 \quad \mbox{and} \quad 
 \partial^{\mu}u\partial_{\mu}u=0. 
\end{equation}
Then we have the local integrability conditions
$ F_{\mu \nu}=0 \quad \mbox{and} \quad D_{\mu}\tilde{B}^{\mu (j)}=0 $. 
The conserved currents are 
\begin{equation}
 J_{\mu}^{(j)} \equiv W^{-1} \tilde{B}_{\mu}^{(j)} W
   =\sum_{m=-j}^{j} J_{\mu}^{(j,m)}P_m^{(j)}. 
\end{equation}
Similarly to Prop \ref{prop:2.1}, 
\begin{prop}we have \\
(a) for $j \geq m \geq 1$, 
 \begin{eqnarray}
   J_{\mu}^{(j,m)} &=&
    \sqrt{\frac{(j+m)!}{j(j+1)(j-m)!}}
    \frac{(-iu)^{m-1}}{(1-|u|^2)^{j+1}} \nonumber\\
  && \hspace{10mm} \times
   \left(  
    \sum_{n=0}^{j-m} \tilde{\alpha} _n^{(j,m)} |u|^{2n} \partial_{\mu}u -
    \sum_{n=0}^{j-m} \tilde{\alpha} _{j-m-n}^{(j,m)} |u|^{2n} u^2 \partial_{\mu}\bar{u}
   \right) ,
 \end{eqnarray}
(b) for $m=0$, 
 \begin{equation}
   J_{\mu}^{(j,0)}=
     i \sqrt{j(j+1)}
    \frac{(\bar{u} \partial_{\mu} u - u \partial_{\mu} \bar{u})}
         {(1-|u|^2)^{j+1}} 
    \sum_{n=0}^{j-1} \tilde{\gamma}_n^{(j,0)} |u|^{2n} , 
 \end{equation}
(c) for $j \geq m \geq 1$, 
\begin{equation}
  J_{\mu}^{(j,-m)}=J_{\mu}^{(j,m)^{\dag}}, 
\end{equation}
where coefficients are
\begin{eqnarray}
 \tilde{\alpha} _n^{(j,m)} &=&
   \frac{n!}{(m+n-1)!} 
    \left(
     \begin{array}{c}
      j-m \\
       n  
     \end{array}
    \right)
    \left(
     \begin{array}{c}
       j+1 \\
        n  
     \end{array}
    \right) , \\
   \tilde{\gamma}_n^{(j,0)} &=&
   \frac{1}{j} 
    \left(
     \begin{array}{c}
       j \\
       n  
     \end{array}
    \right)
    \left(
     \begin{array}{c}
        j \\
       n+1  
     \end{array}
    \right) .
\end{eqnarray}
\end{prop}
 \section{Discussion}
In this paper, we gave explicit forms to nontrivial conserved currents of submodels of $CP^1$-model, $CP^1$-like models and $QP^1$-model in $(1+2)$-dimensions.
Our result is a generalization of that of \cite{A-F-G}. 

We also take an interest in whether or not our results can be generalized for the $CP^N$-model or $QP^N$-model in $(1+2)$-dimensions. 

On the other hand, they in \cite{A-F-G} defined a four dimensional integrability and applied their method to the self-dual Yang-Mills model or other models. But their arguments are a bit unclear (at least to us). 

In a forthcoming paper, we shall make an attack to these problems. 

 \section*{Acknowledgements}
Kazuyuki Fujii was partially supported by Grant-in-Aid for Scientific Research (C) No. 09640210. Tatsuo Suzuki is very grateful to Tosiaki Kori for valuable discussions. 

\end{document}